\documentclass[11pt]{article}
\usepackage[a4paper,margin=2.4cm]{geometry}
\usepackage[T1]{fontenc}
\usepackage{mathptmx}
\usepackage{helvet}
\usepackage{graphicx,amsmath,xcolor,float,setspace,titlesec,caption}
\usepackage[super,comma,sort&compress]{natbib}
\usepackage{tikz}
\usepackage{tcolorbox}
\providecolor{boxframe}{HTML}{0B6E71}
\providecolor{boxback}{HTML}{F3F7F7}
\newtcolorbox{readerbox}[1]{colback=boxback, colframe=boxframe, coltitle=white, fonttitle=\sffamily\bfseries\small,
  title=#1, arc=2pt, boxrule=0.6pt, left=7pt, right=7pt, top=5pt, bottom=5pt, before skip=10pt, after skip=10pt,
  fontupper=\small}
\newcommand{\boxterm}[1]{\textbf{#1}}

\newfloat{readerboxfloat}{tbp}{lob}
\newenvironment{floatingbox}[1]{\begin{readerboxfloat}[tb]\begin{readerbox}{#1}}{\end{readerbox}\end{readerboxfloat}}

\usepackage[hidelinks]{hyperref}

\definecolor{ncteal}{HTML}{0B6E71}
\titleformat{\section}{\sffamily\bfseries\large\color{ncteal}}{}{0pt}{}[\vspace{1pt}\color{ncteal}\titlerule\vspace{2pt}]
\titlespacing{\section}{0pt}{14pt}{6pt}
\titleformat{\subsection}{\sffamily\bfseries\normalsize}{}{0pt}{}
\titlespacing{\subsection}{0pt}{10pt}{4pt}
\title{\sffamily\bfseries\Large Learning a non-linguistic code for inferred rules from reward}
\author{\normalfont Cristiano Capone\\[2pt]\normalsize Quantitative Health Models and Bioinspired Technologies Lab, Istituto Superiore di Sanit\`a, Rome, Italy\\[1pt]\small\texttt{cristiano.capone@iss.it}}
\date{}
\begin{document}
\maketitle
\vspace{-8pt}{\color{ncteal}\hrule height 1pt}\vspace{4pt}
\section*{Abstract}
\noindent 
How can a rule inferred from examples reach someone who never saw them, without a shared code? Patients with
severe aphasia do it by gesture or sketch. One network sees worked examples and emits eight invented symbols; a second,
blind to them, applies them to a new input. Rewarded for the second's success, the first learns a code carrying
rules to three-step transformations training never presents, which new learners acquire. Like invented human
languages, the code has two regimes: under reward alone the speaker drifts to one message, as human languages lose
words under plain transmission; expressive pressure keeps messages differentiated. Success on new rules tracks how
much the message says about the rule, not how varied messages are. The learning signal shapes the code: reward
sorts many rules under few fixed labels; the listener's error gradient gives each rule a region of similar
messages, telling rules apart far better.

\section*{Introduction}

People pass on rules they have inferred, even without language. The brain network that supports language is
sharply delineated and responds to little else \cite{shain2026,fedorenko2024nrn}; it stays quiet during logical
reasoning, and people who have lost it can still do algebra and solve formal problems
\cite{monti2012,varley2005,fedorenko2024}. Most directly, Kean \emph{et al.}\cite{kean2026} gave two people with
severe aphasia examples of a hidden rule mapping one list of numbers onto another: they worked it out at the level
of healthy controls and conveyed it by gesture or drawing. Whatever carried that rule, it was not English.

\begin{figure}[tb]
\centering
\includegraphics[width=\textwidth]{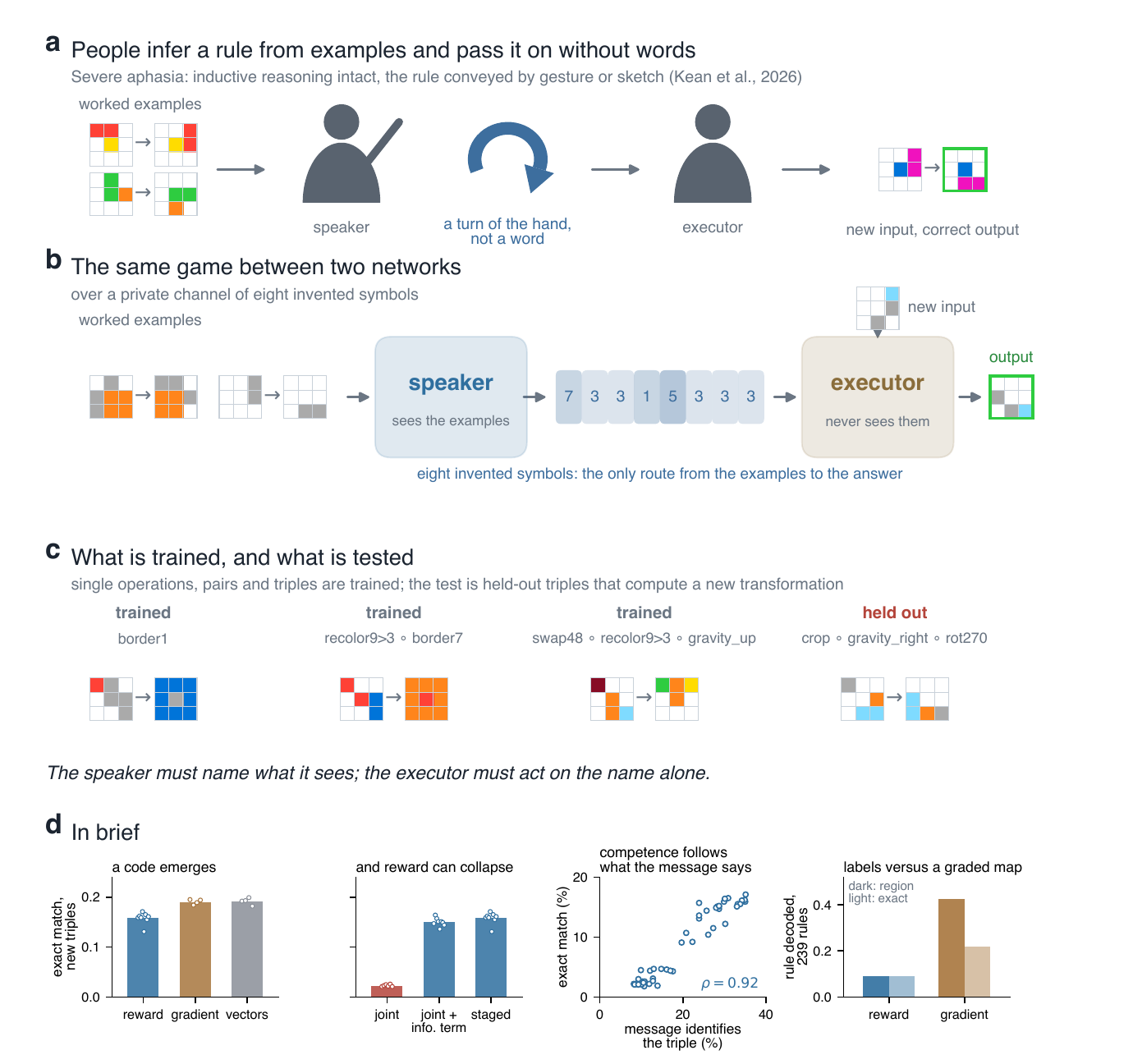}
\caption{\textbf{The task, and the human ability it models.}
\textbf{(a)} People with severe aphasia infer a rule from examples and convey it without words \cite{kean2026}; the
speaker and the executor of the game are these two roles.
\textbf{(b)} The game has the same shape: the speaker sees the examples and sends eight invented symbols, the
executor sees only a new input and those symbols. \textbf{(c)} Example episodes; generalisation is tested on
held-out triples training never computes, even after its rotations, reflections and colour changes.
\textbf{(d)} In brief: exact match on new triples for the three
channels; the same under joint, joint with the information term, and staged experience; that accuracy against how
often the message identifies the rule, over 50 reward-trained runs; and how often a rule is decoded from the
message among 239 families, by nearest region and by exact message.}
\label{fig:paradigm}
\end{figure}

This tells us such a code exists, not what one looks like or what it takes to build one. Those are computational
questions, and we ask them in a game with the same shape as the human task (Fig.~\ref{fig:paradigm}). A
\emph{speaker} network sees a few worked examples of a transformation and emits a short string of discrete
symbols; an \emph{executor} network receives the symbols and a fresh input, and must produce the right output.
The executor never sees the examples, so everything it knows about the rule travels through the symbols, and
nothing about the symbols is given in advance: no symbol has an assigned meaning, there is no grammar, and nothing
supervises what any symbol should stand for.

The idea that thought runs on a structured, symbol-like code distinct from any spoken language is old in
cognitive science \cite{fodor1975,piantadosi2016,quiltydunn2023,dehaene2022}, and people who must communicate
without a shared language invent such codes quickly, by drawing or gesture \cite{galantucci2005,fay2010,motamedi2019}.
Invented codes have been studied formally since the signalling games of Lewis \cite{lewis1969}, which have pooling
equilibria in which states share a signal, reached by reinforcement learners with positive probability
\cite{skyrms2010,huttegger2010,barrett2009}. Emergent-communication research has taken such games to neural agents
\cite{lazaridou2020}, almost always as referential games in which a listener picks an object out of a set
\cite{lazaridou2017,havrylov2017,chaabouni2019,chaabouni2020,ren2020,li2019}, or communicates a concept shared by a
set of objects \cite{mu2021}. It has shown that gradients through the channel learn a protocol more reliably than
reward \cite{foerster2016}, that rewarded codes need not be compositional \cite{kottur2017} and carry little more
than the task demands \cite{kharitonov2020}, that success at the game is a coarse measure of what a code carries
\cite{lowe2019}, and that the objective mixes information with co-adaptation between speaker and listener
\cite{rita2022}. Here the message is not a label to be matched but a rule to be executed on a new input: a model of rule
transmission rather than of reference.

We report five findings, in the order of the Results. \emph{First}, a code emerges: it carries rules to
three-step transformations that training never computes, and a new learner can acquire it. Establishing this
required an audit of the held-out split: training episodes are rotated, reflected and recoloured, so almost every
held-out pair of operations is a transformation training already presents, and only held-out triples test
composition (Supplementary Section~S1). \emph{Second}, discreteness costs little, as long as perception is learned
outside the channel. \emph{Third}, reward and gradient build different kinds of code: reward a few labels, each
shared by many rules; gradient a graded map in which each rule occupies its own region. \emph{Fourth},
reward-trained communication has a measurable degenerate solution, the same message whatever the rule; staged
experience or a pressure against uninformative messages prevents it, and across the manipulations we tried,
competence goes with how much the message says about the rule. Human learners who pass on an invented language
show similar regimes. \emph{Fifth}, reward-driven naming saturates as rules are added, and neither of the two
capacity increases we tried lifted it.

\section*{Results}
\subsection*{A code emerges, and it composes}

A discrete code appears, and it carries the rule. A held-out combination tests composition only if training never
computes the same transformation, and training offers more than its list of programs: geometric operations compose
into one another, colour operations commute with spatial ones, a held-out pair can equal a trained triple whose
last step changes nothing, and every training episode is rotated or reflected with its colours permuted, so each
trained program is also trained in disguise. Checking every held-out program against every training program under
all of these transforms (Methods), only 6 of the 212 held-out pairs compute a new transformation, while 100 of the
300 held-out triples do, so we test generalisation on the triples (Supplementary Section~S1).

\begin{floatingbox}{{Box 1 $|$ The terms used throughout}}
\boxterm{Speaker and executor.} The speaker sees a few worked examples of a transformation and writes eight symbols
from an alphabet of thirty-two. The executor sees only those symbols and a new input grid, and must write the
output. Nothing about the symbols is given in advance.\\[3pt]
\boxterm{Three channels.} \emph{Reward}: the speaker learns only from whether the executor succeeded. \emph{Gradient}:
the executor's error signal passes back through the symbols. \emph{Continuous}: a real-valued vector replaces the
symbols; it is the reference for what a channel can carry.\\[3pt]
\boxterm{Order of experience.} \emph{Staged}: single operations first, then pairs, then triples. \emph{Joint}: all of
them from the first step. \emph{Reversed}: pairs first, then single operations, then triples.\\[2pt]
\begin{center}\begin{tikzpicture}[x=1.55mm, y=1mm, font=\scriptsize, every node/.style={inner sep=1pt}]
  \node[anchor=east] at (-1,7) {staged};
  \fill[boxframe!25] (0,5) rectangle (20,9); \fill[boxframe!50] (20,5) rectangle (50,9); \fill[boxframe!80] (50,5) rectangle (80,9);
  \node at (10,7) {singles}; \node at (35,7) {pairs}; \node[white] at (65,7) {triples};
  \node[anchor=east] at (-1,1) {joint};
  \fill[boxframe!60] (0,-1) rectangle (80,3); \node[white] at (40,1) {singles, pairs and triples together};
  \draw[->,gray] (0,-3) -- (80,-3); \node[anchor=north,gray] at (0,-3.5) {0}; \node[anchor=north,gray] at (80,-3.5) {32\,000 steps};
\end{tikzpicture}\end{center}
\boxterm{New triples.} Held-out three-step transformations that no training program computes, even after the
rotations, reflections and colour changes used to augment training: 100 of the 300 held out. They are the test of
composition. Accuracy on them is \emph{exact match} of the whole output grid; copying the input scores zero.\\[3pt]
\boxterm{Constant-message control.} The same executor with every message replaced by one fixed message: what it
achieves without the channel.\\[3pt]
\boxterm{Names and decoding.} A rule's \emph{name} is its most frequent message; counting distinct names says how
many rules share one. \emph{Decoding} asks how often a message identifies its rule on episodes the decoder never
saw (chance 1\% for 100 triples).\\[3pt]
\boxterm{Collapse.} The degenerate solution in which the speaker sends the same message whatever it sees; measured
as distinct messages per held-out rule.
\end{floatingbox}

\begin{figure}[tb]
\centering
\includegraphics[width=\textwidth]{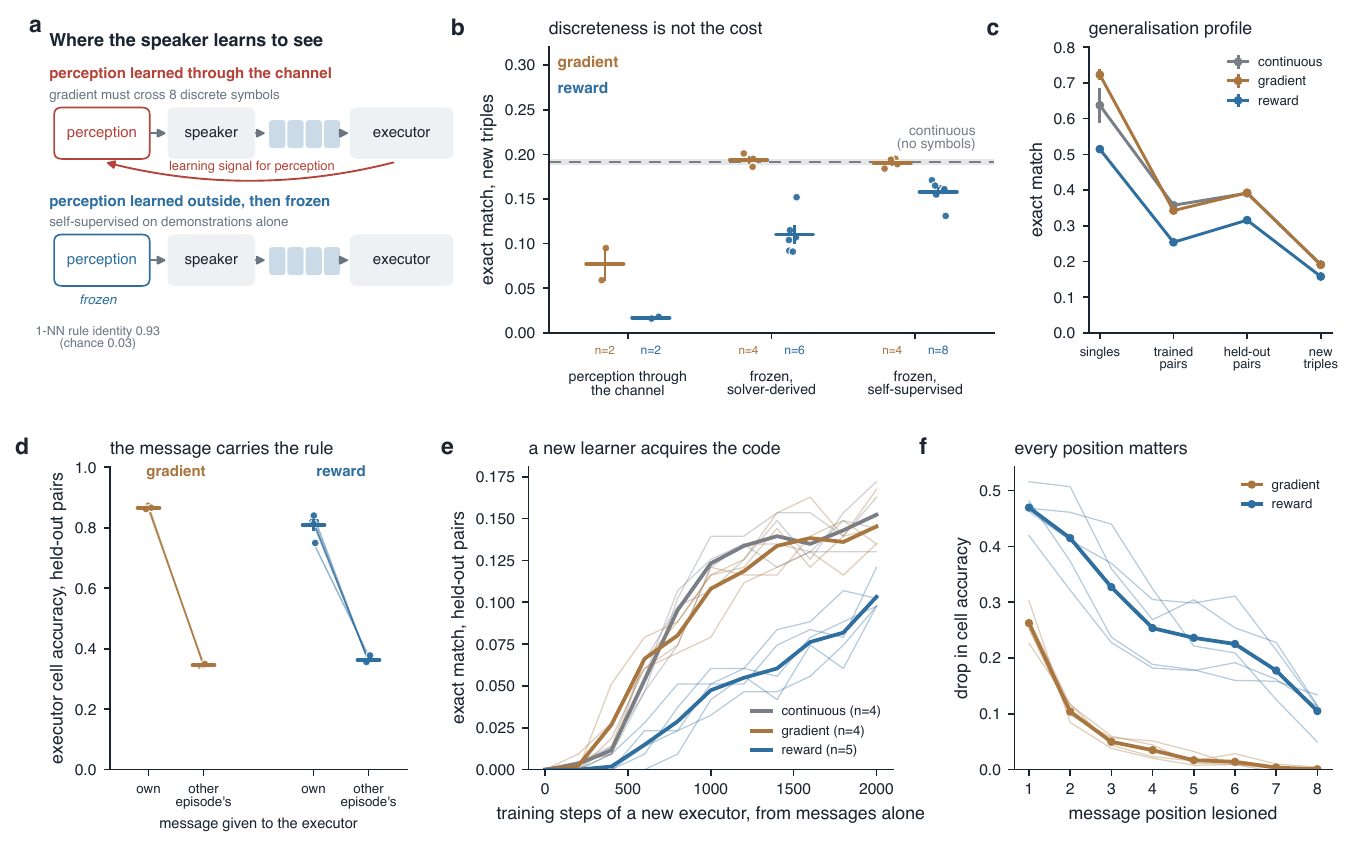}
\caption{\textbf{A discrete code emerges and composes.}
\textbf{(a)} Two ways for the speaker to learn to see: through the channel, or outside it and then frozen. The
frozen encoder identifies the rule of an unseen episode in 93\% of cases, against 3\% by chance. \textbf{(b)}
Exact match on new triples, gradient code (ochre) and reward code (blue), real-valued channel as a grey band;
points are runs. \textbf{(c)} The same channels across splits. \textbf{(d)} Executor accuracy with its own
message and with another episode's. \textbf{(e)} A fresh executor learning from the frozen speaker's messages.
\textbf{(f)} Drop in accuracy when one position is replaced by a random symbol.}
\label{fig:emergence}
\end{figure}

Differences between conditions come with 95\% confidence intervals from a hierarchical bootstrap over runs and
held-out programs (Methods; Supplementary Table~S11 lists every contrast). With perception learned outside the
channel (next section), a speaker restricted to eight symbols from a vocabulary of thirty-two lets the executor
reproduce a new triple in 19.1\% of episodes under gradient training and in 15.8\% under reward alone, against
19.2\% for a channel that passes real-valued vectors (Fig.~\ref{fig:emergence}b). The three channels share the same profile (Fig.~\ref{fig:emergence}c): single
operations are easiest, pairs harder, new triples hardest.

Three tests show that the symbols do the work. Given the message the speaker had written for another episode,
the executor's cell accuracy on held-out pairs fell from 0.87 to 0.35 under gradient training and from 0.81 to 0.36
under reward, and exact match to 0.01 (Fig.~\ref{fig:emergence}d). A fresh executor that sees only the frozen
speaker's messages learns to act on them within two thousand steps (Fig.~\ref{fig:emergence}e), and after eight
thousand solves 11 to 12\% of new triples from the reward code and 15\% from the gradient code, a little below the
original executors, against 0.4\% when the messages are shuffled between episodes (Supplementary Section~S5): the
code is public, and a new learner can acquire it. And replacing one symbol at a time shows the reward code
spreading information along the message, the drop in cell accuracy falling from 0.47 at the first position to 0.11
at the last, while the gradient code concentrates it in the first two positions (Fig.~\ref{fig:emergence}f).

\subsection*{Discreteness costs little; learning to see through the channel costs a great deal}

Trained end to end, the discrete channel looks far worse than the real-valued one: on new triples the gradient
code reaches 7.7\% and the reward code 1.7\% (two runs each). This is a fact about how perception is learned, not
about symbols. We trained the speaker's visual front end separately, with a contrastive objective on the
demonstrations that never sees the query's answer, and froze it. With perception copied from a trained real-valued
model, the two codes reach 19.4\% and 11.0\%; with perception trained by self-supervision on the demonstrations
alone, without the query's answer or any label of the rule, 19.1\% and 15.8\%, the latter with a standard error of
0.4 point over eight runs, against 19.2\% for the real-valued channel (differences $-0.1$ points, 95\% confidence
interval $-1.5$ to 1.4, and $-3.4$ points, $-5.6$ to $-1.4$; Fig.~\ref{fig:emergence}b). On single operations the
gradient code reaches 72.3\%, above the real-valued channel at 63.8\% (Fig.~\ref{fig:emergence}c). What
discreteness costs is not accuracy but a learning signal for perception; once perception has its own signal,
eight symbols carry what an unconstrained vector carries under gradient training, and most of it under reward.

\subsection*{Reward and gradient grow different kinds of code}

\begin{figure}[tb]
\centering
\includegraphics[width=\textwidth]{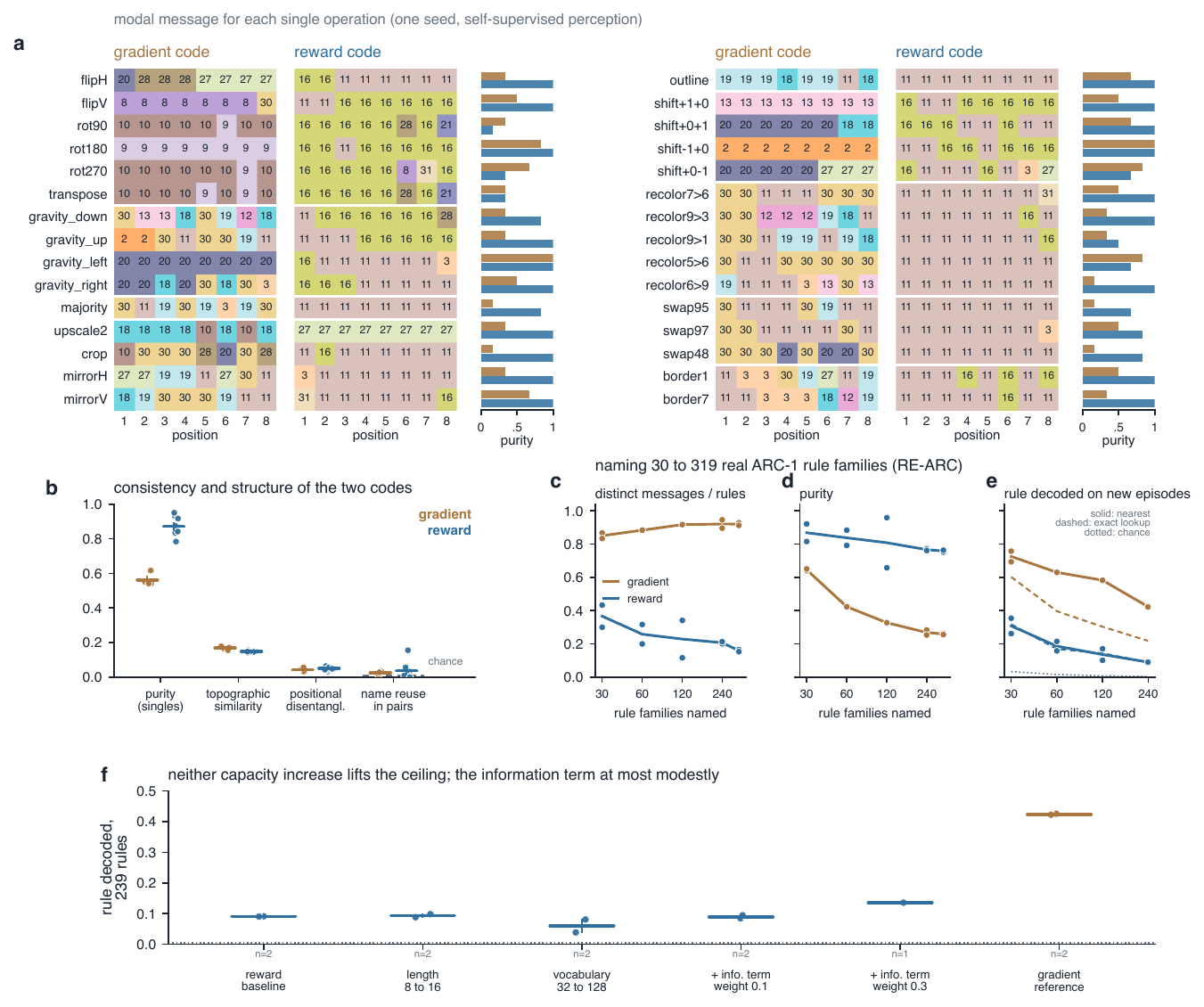}
\caption{\textbf{Reward and gradient grow different kinds of code.}
\textbf{(a)} The most frequent message for each of the thirty single operations, one run per channel; colour and
number give the symbol at each position, bars give purity. \textbf{(b)} Purity, topographic similarity, positional
disentanglement, and how often a pair's message reuses a component's name (dashed line, chance). \textbf{(c--e)}
Naming 30 to 319 rule families from ARC-1: messages per family, purity, and how often the family is decoded from
the message, by nearest neighbour (solid) and exact lookup (dashed), with chance (dotted). \textbf{(f)} At 239
families, decoding after doubling the message length, quadrupling the vocabulary, or adding the information
term.}
\label{fig:twocodes}
\end{figure}

The two channels reach similar accuracy with codes that look different (Fig.~\ref{fig:twocodes}a). The reward code
uses few symbols in fixed positions: in the run shown, one symbol marks nearly every recolouring, swap and border,
another the geometric operations and a third gravity, with a second symbol telling members of a family apart. The
gradient code gives most operations a message of their own, drawn from many more symbols, and is less consistent:
an operation receives its most frequent message in 87\% of episodes under reward and in 56\% under gradient
(Fig.~\ref{fig:twocodes}b). Neither code names an operation the same way inside a pair as alone: topographic
similarity is low for both, positional disentanglement is near zero, and a pair's message contains a component's
name in 4\% of cases under reward and 2\% under gradient, against about 1\% by chance. Both compose by some other
route.

The same contrast holds on real rules, and we can state it by asking whether a message identifies its rule on
episodes the decoder has not seen. We trained both channels to name between 30 and 319 rule families of the ARC-1
training set, regenerated at easy difficulty, and decoded the family from the message on one half of the episodes
after fitting on the other (Fig.~\ref{fig:twocodes}c--e). The gradient code identifies the family in 69 to 76\% of
cases at 30 families and in 42\% at 239, against chance levels of 3.3\% and 0.4\%; the reward code in 26 to 35\%
and 9\% (difference at 239 families 33.4 points, 30.2 to 36.6). The two decoders then separate the codes. Exact
lookup succeeds only when the same message recurs; nearest neighbour in Hamming distance succeeds when a family's
messages occupy a consistent region. For the reward code they agree, 9\% and 9\%: it uses few labels, consistently,
48 to 51 messages for 239 families with purity 0.76. For the gradient code nearest neighbour is twice exact lookup,
42\% against 22\% (difference 20.5 points, 17.5 to 23.4): the same family rarely receives the identical message,
purity falls to 0.25, but its messages cluster in one region. The reward code behaves as a set of labels and the
gradient code as a graded map; a mutual information estimated in the sample would have hidden this
(Supplementary Section~S2).

\subsection*{Staged experience keeps reward training out of a degenerate solution}\label{sec:collapse}

Reward-trained codes composed only when the operations arrived in a staged order. On new triples, staged
experience reached 15.8\% (eight runs), reversed staging 3.2\%, and joint experience, with all operations from the
first step, 2.3\% (six runs; staged minus joint, 13.5 percentage points, 95\% confidence interval 9.1 to 18.4;
exact permutation test on run means, $p = 0.0003$). Gradient-trained codes were unaffected, at 18.4 to 19.3\% in
all three orders (interaction between order and channel, 13.8 points, 9.0 to 18.7; Fig.~\ref{fig:collapse}b). Such a dissociation asks for a mechanism.

We first asked whether the difference lay in credit assignment, since a scalar reward can only reinforce a message
as a whole. Giving each symbol its own credit, from the gain it produced in the executor's state, left joint
training where it was. Two further accounts also failed: more episodes with an operation alone, and more samples per episode.

The messages themselves explain the dissociation. Under joint training the reward-trained speaker sent nearly the same
message whatever it saw (Fig.~\ref{fig:collapse}c). This degenerate solution is the extreme case of a pooling
equilibrium \cite{skyrms2010,huttegger2010}, and under the task reward alone it sustains itself: once the message
carries no information, the executor learns to ignore it, and no variation in the message is rewarded. It also
explains why the staged order helps: in the first stage, with a few dozen equally likely single operations and
nothing else, a constant message is measurably worse than a discriminative one, so the degenerate solution is
never entered.

If that is the whole story, removing the degeneracy directly should replace the curriculum. A term in the
speaker's objective that favours messages depending on what it sees \cite{eccles2019} does exactly that: joint
training then reaches 15.0\% of new triples (eight runs; against joint training without it, 12.7 points, 8.4 to
17.5; permutation $p = 0.0003$), not detectably different from staged experience ($-0.8$ points, $-3.0$ to 1.2),
with no staging at all, and the diagnostic moves with it (Fig.~\ref{fig:collapse}c,d). The gradient channel never
enters the degenerate solution, because its estimator carries information about what the executor needs, and it is
correspondingly indifferent to training order.

\subsection*{Competence goes with how much the message says about the rule}

What the information term restores is information about the rule, and competence goes with it. Counting distinct messages
overstates how differentiated a code is, because a speaker can vary its message between episodes while giving many
rules the same name; we measured instead how often the message identifies its rule on held-out episodes (Methods).
Manipulations that add variety without much information restore little. A reward that pays only for the cells the
rule changes doubles the number of names for the thirty operations, 13 to 16 against 6 or 7, but half of them still
share one name; its messages identify a new triple in 10 to 11\% of episodes, against 8 to 10\% for the collapsed
code and 24 to 35\% for the staged one, and it solves 1.9\% of new triples. Keeping the speaker exploring for longer
adds a little of both (Supplementary Section~S5).

Across the 50 reward-trained runs on this library, in twelve conditions that vary the order of experience, the
reward, the information term, exploration, credit assignment and perception, accuracy on new triples rises with how
often the message identifies the triple (Spearman $\rho = 0.92$, 95\% confidence interval 0.84 to 0.96;
Fig.~\ref{fig:collapse}d; 0.89 with six runs whose policy a stability pressure kept soft). The gradient code
identifies the triple most often, 47 to 56\%, and is the most accurate. This is a relation across manipulations,
not a demonstration that information is all a failing code lacks: the soft-policy runs fall furthest below it,
their most probable messages as informative as a staged code's while their executors, trained on noisy samples of
them, solve 6.5 to 9.1\% (Supplementary Section~S5).

A new executor tests the same point from the listener's side, following the decomposition of the objective into
information and co-adaptation \cite{rita2022}. Trained for eight thousand steps on a frozen speaker's most probable
messages, it solves 11 to 12\% of new triples from staged and information-term codes, at most 3\% from joint,
changed-cell and slower-exploration codes, and 0.4 to 0.5\% from shuffled messages (Supplementary Table~S7). With
the architecture and budget we tested, a fresh listener does not recover competence from a failing code, except in
part where the policy had been kept soft (Supplementary Section~S5).

\subsection*{The degenerate solution beyond these networks and this library}

The same failure and the same remedy appear in a pretrained language model. In a pilot, two copies of
Qwen2.5-0.5B, adapted with low-rank adapters, played the game with the same reward recipe. Joint and staged
experience both collapse the code in all four seeds each; the information term gives an informative code in seven
of eight evaluated checkpoints, which solve 6 to 8\% of new triples against a 1.3\% control (Supplementary
Section~S9). The pilot is small and its reward differs from that of the main experiments; it suggests, without
establishing, that the degenerate solution and its remedy are not peculiar to the small networks here.

Whether a family of rules enters the degenerate solution, however, we could not predict. In six more libraries (Fig.~\ref{fig:collapse}e;
four subsets of the original, a second library of seventeen operations sharing none with it, and that library with
six near-duplicates), every subset of the first library collapses under joint experience except a random
seventeen, while the second library never collapses, staged and joint experience landing within a point of each
other and of the real-valued channel (Supplementary Section~S10). Four candidate predictors failed: the number of
operations, perceptual confusability (Fig.~\ref{fig:collapse}f), the partial credit one operation earns for
another's output, and the value of the message to the executor. The claim is therefore not that reward-trained
composition needs staged experience, nor a rule about which families are at risk, but that reward-trained
communication has a degenerate solution, that a measurable quantity says when a family has entered it, and that
staged experience or an explicit pressure removes it where it occurs and costs nothing where it does not.

\subsection*{People show similar regimes}

\begin{figure}[tb]
\centering
\includegraphics[width=\textwidth]{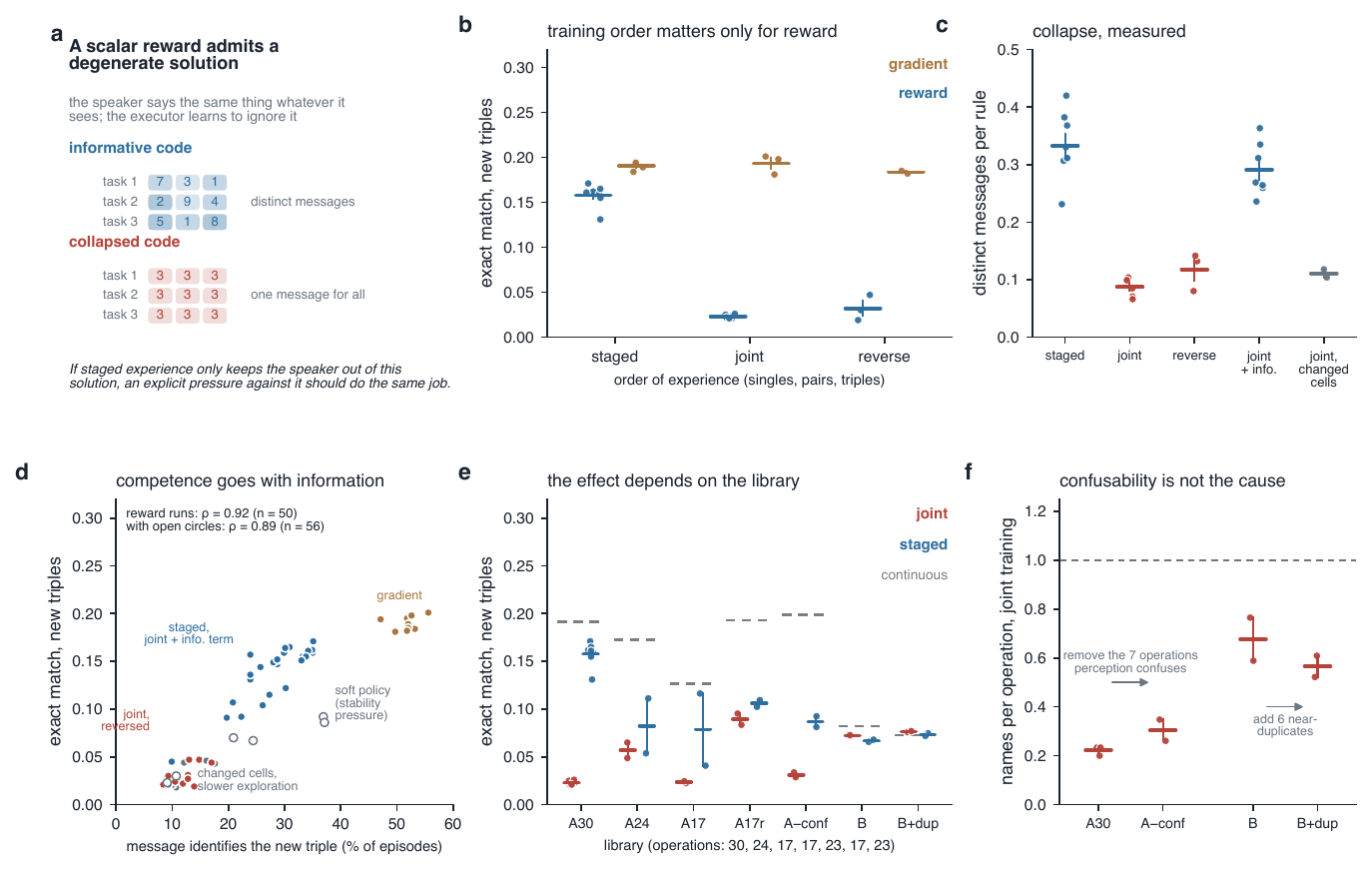}
\caption{\textbf{Staged experience, collapse, and what an information term restores.}
\textbf{(a)} The hypothesis. \textbf{(b)} Exact match on new triples after staged, joint and reversed experience,
both channels. \textbf{(c)} The diagnostic: distinct messages per held-out rule, reward channel. \textbf{(d)}
Exact match against how often the message identifies the triple, one point per run: staged experience or the
information term (blue), joint or reversed experience (red), changed-cell reward or slower exploration (grey), a
policy kept soft (open), gradient channel (ochre). \textbf{(e)} Joint against staged reward training in seven
libraries. \textbf{(f)} Names per operation under joint training, before and after two manipulations of
confusability.}
\label{fig:collapse}
\end{figure}

\begin{figure}[tb]
\centering
\includegraphics[width=\textwidth]{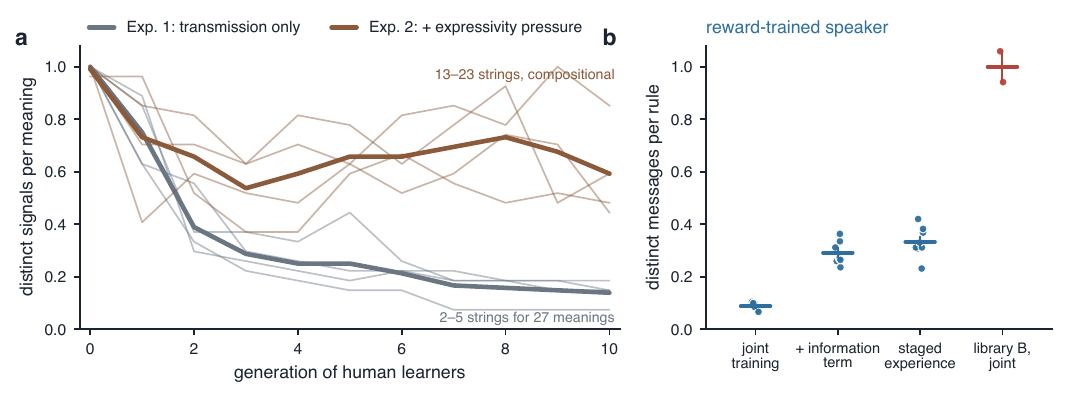}
\caption{\textbf{People show similar regimes.} \textbf{(a)} Human iterated learning \cite{kirby2008}:
twenty-seven meanings passed across ten generations, with and without a pressure against homonyms. Thin lines are
chains, thick lines their mean. \textbf{(b)} The reward-trained speaker under the conditions of
Fig.~\ref{fig:collapse}, measured with the same statistic; one point per run.}
\label{fig:human}
\end{figure}

In the iterated-learning experiment of Kirby, Cornish and Smith
\cite{kirby2008}, each participant learns an artificial language for twenty-seven meanings from the previous
participant's output. Re-analysed with the statistic of Fig.~\ref{fig:collapse}c, the language collapses under plain
transmission from twenty-seven distinct signals to between two and five within ten generations, 0.14 per meaning,
and keeps 0.59 signals per meaning when homonyms are filtered out of each training set, a pressure for
expressivity (difference at the tenth generation 0.45, 95\% confidence interval 0.17 to 0.74; Fig.~\ref{fig:human}).
The pressures differ, and so do the mechanisms; what is shared is that a code shaped only by the pull towards
compression degenerates, and a pressure for expressivity keeps it differentiated, in human learners and in our
speaker, measured the same way (Supplementary Section~S10).

\subsection*{Reward-driven naming saturates, and the two capacity increases we tried did not lift it}

If reward-driven naming saturates because the speaker cannot fit more names into eight positions, more capacity
should help. Neither of the two increases we tried did (Fig.~\ref{fig:twocodes}f; two runs per condition unless
stated). At 239 families, nearest-neighbour decoding of the reward code is 9\% at baseline. Doubling the message
length to sixteen symbols adds distinct messages, from 48 to 51 up to 63, and no decodable information, 9\% and
10\%. Quadrupling the vocabulary to 128 symbols lowers it, to 4\% and 8\%. The term that penalises constant
messages does not lift this ceiling at the weight used above, 9\% and 10\%, and lifts it modestly at a higher
weight, 14\% from one run, against 42\% for the gradient code. Within the range we tested, saturation is a separate property of
reward-trained naming, neither a capacity limit nor collapse at a larger scale.

\section*{Discussion}

We set out to build the smallest system in which one can ask what a code for transmitting an inferred rule looks
like when nothing about it is given in advance, and to ask it with reward alone, because no error gradient crosses
from one person to another. A code appears, it carries the rule to three-step combinations that training never
presents, even through its augmentation, and to a new learner, and it is cheap once perception has a learning signal of its own. Its structure depends on how it was learned: reward builds a small set of labels into which many
rules are sorted, gradient a graded map in which each rule occupies its own region.

Learning from reward is fragile in a specific way. A scalar reward admits a degenerate solution in which the
speaker says the same thing whatever it sees, and joint training falls into it. Neither the degeneracy nor its
remedy is new: reinforcement learners settle in pooling equilibria in signalling games with a handful of states
\cite{huttegger2010,barrett2009}, and the pressure that removes it here is the speaker bias of Eccles
\emph{et al.}\cite{eccles2019}. What this system adds is the setting, a rule inferred from examples and executed
by someone who never saw them, and three measurements in it: the degenerate solution is entered or avoided
according to the order of experience, it goes with the failure to transmit new transformations, and competence
goes with how much of the rule the message carries. The same degeneracy and remedy appear in a pilot with a
pretrained language model (Supplementary Section~S9), and the languages people invent in the laboratory degenerate
under plain transmission and stay differentiated under a pressure for expressivity \cite{kirby2008}, a contrast
that theories of cultural language evolution place at the origin of linguistic structure \cite{kirby2015}. The
pressures and the mechanisms differ, and we do not claim they are the same; what the parallel shows is that the
regime a code ends up in is set by a pressure on the code, in human learners and in our speaker, measured the same
way.

Two results bound that account. First, what counts is the information the code carries, not its variety:
manipulations that multiplied distinct messages without adding much information restored little competence, and a
new executor trained on a frozen speaker's messages learns informative codes and does not recover competence from
uninformative ones within the budget we tested. These are relations across manipulations and a test with one kind
of listener; they narrow the explanation to missing information without excluding every other difficulty of
learning. Second, the saturation of reward-driven naming at scale is a different phenomenon: neither of the two
capacity increases we tried lifted it, and the term that prevents collapse lifts it at most modestly. Which task
families are at risk of collapse is likewise not predicted by any quantity we tried.

Four limitations bound the claim. Composition is tested on held-out triples only, because the augmentation of
training covers all but six of the held-out pairs, and only 100 of 300 held-out triples remain new. We found no
evidence that the executor works serially. Transfer to the original ARC-1 problems is marginal, three of eighty
held-out families at easy difficulty and none at the original difficulty, so the system is a model of rule
transmission rather than a solver. And the categories the reward code forms bear little relation to how people
describe the same rules in English \cite{acquaviva2022} (Supplementary Section~S7): consistent with the premise,
since English is the channel the patients did not use, but the human comparison available today is at the level of
mechanism, not of content. The comparison that
would speak to content, people passing inferred rules through a restricted non-linguistic channel under conditions
that vary the pressure on the code, does not yet exist, and the diagnostics developed here are ready for it.

\section*{Methods}
\subsection*{Tasks}
\textbf{Compositional library.} Library A holds thirty deterministic grid-to-grid operations in five
families: geometry (four rotations, two reflections, transpose), gravity (four directions), translation
(four one-cell shifts), colour (five recolourings, three colour swaps, majority fill) and size and
shape (twofold upscale, bounding-box crop, horizontal and vertical mirror-concatenation, outline, two
borders). A rule is a sequence of one to three operations applied left to right; sequences that act as
the identity on five of six probe grids are discarded. Input grids are drawn per episode, three to six
cells a side, in one of three styles (sparse noise, one to three solid rectangles, rectangles plus
noise) with two to five of the nine non-background ARC colours, and padded to a square canvas of
twelve with a dedicated padding class, so that the output shape is part of what must be predicted.
Library B holds seventeen operations from disjoint families (counting, symmetry, connectivity,
ordering, masking) and is used with the same generator and splits.

\textbf{Compositional split.} All single operations are training rules. Ordered pairs are split into
639 training and 212 held-out combinations (204 and 68 for library B): both components of a held-out
pair appear in training, the pair never does. Self-compositions are split 11 training and 5 held out;
triples 600 and 300. The splits are fixed by a seed independent of the model seed.

\textbf{Real ARC-1 rules.} The RE-ARC generators \cite{hodel2024} provide one procedural generator per
task of the ARC-1 training set. We use all 400, split into 320 training and 80 held-out rule families,
sampled at the easy end of each generator's difficulty range (0 to 0.15) on a canvas of sixteen, so
that the task is within reach of a four-million-parameter model. For the naming experiments the
training set is restricted to the first 30, 60, 120 or 240 families. The ARC-1 evaluation set and
ConceptARC are monitored and never trained on.

\subsection*{Networks}
\textbf{Speaker.} Each demonstration pair is embedded by a shared token embedding and a learned
positional map, mixed by a $1\times1$ convolution, and encoded by two residual blocks of $3\times3$
convolution, row attention, column attention and feed-forward layers; the result is downsampled twice
and pooled by a learned query into one vector per pair. A two-layer transformer over the pairs,
followed by masked mean pooling, gives a single 128-dimensional context. We call the pair encoder and
this transformer the perception module (2.22\,M parameters). A gated recurrent unit decodes the
context into $L=8$ symbols from a vocabulary of $V=32$, consuming its own previous symbol. Speaker
total: 2.34\,M parameters.

\textbf{Executor.} The query grid is embedded on the same canvas and updated once per symbol by two
blocks of the same form, conditioned by feature-wise affine modulation on the symbol's executor-side
embedding and a step embedding; a $1\times1$ convolution reads out eleven classes per cell. The
executor never sees the demonstrations (1.41\,M parameters). Speaker and executor share no parameters
and no symbol embedding: only the integers cross the channel.

\subsection*{Channels}
\textbf{Reward.} Symbols are sampled and the speaker receives no gradient from the executor. The
policy is trained by REINFORCE with a group baseline: four messages are sampled per episode and the
advantage is the within-group standardised return. Entropy is held above a floor of 0.3\,nat by a
hinge penalty, with no entropy bonus; the speaker's learning rate is 0.3 times the executor's. The
executor is trained by cross-entropy on the same samples. \textbf{Gradient.} The same discrete
symbols with a straight-through Gumbel-softmax estimator at temperature 1, so that the executor's loss
reaches the speaker through a biased surrogate gradient. \textbf{Continuous.} No discretisation:
eight vectors of dimension 128. \textbf{Information term.} Where stated, the speaker's loss includes
$-0.1\,[H(\bar p) - \overline{H(p)}]$, where $H(\bar p)$ is the entropy of the symbol distribution averaged over
the episodes of a batch and $\overline{H(p)}$ the mean entropy of each episode's distribution, per position
\cite{eccles2019}; the bracket is positive when messages differ across episodes and zero when they do not. The
term is a differentiable function of the speaker's own output probabilities, added to the policy-gradient loss
rather than to the reward, and it carries nothing from the executor; weight 0.3 where stated.

\subsection*{Training}
AdamW ($\beta=0.9, 0.95$; weight decay 0.01), learning rate $3\times10^{-4}$ with 500 warm-up steps and
cosine decay, batch 32 episodes of two to four demonstrations, gradient-norm clipping at 1, and 32\,000
steps (20\,000 for the RE-ARC runs). Each episode is augmented by one of the eight dihedral transforms
and a random permutation of the non-background colours. Staged experience runs three stages with a
cosine restart at each boundary: 8\,000 steps of single operations; 12\,000 of singles, training pairs
and repeats; 12\,000 adding training triples. Joint experience samples the final mixture from the
first step; reversed experience runs pairs before singles. The three arms of the primary contrasts were run with six to eight
seeds and most other configurations with two to six; a few control conditions have a single run, and the number of runs of every condition is listed in
Supplementary Table~S11. Runs are shown individually in the figures, with mean and standard error where there is
more than one.

\textbf{Perception.} Three sources: learned end to end through the channel; copied from a finished
continuous-channel run and frozen; or trained by self-supervision and frozen. The self-supervised
encoder is trained for 6\,000 steps with an InfoNCE objective (temperature 0.1) whose two views of an
episode are two disjoint halves of its demonstration pairs, the same rule on different grids; it uses
no query answers and no labels. Its quality is the one-nearest-neighbour rule identity of held-out
episodes, 0.925 against a chance level of 0.033.

\subsection*{Novelty of the held-out programs}
A held-out program tests composition only if the transformation it computes is absent from what training
actually presents. Training presents more than the listed training programs: every episode is transformed by one
of the eight dihedral transforms $T$ and a permutation $s$ of the non-background colours, applied to demonstrations
and query alike, so a training program $P$ is trained as the function $sTPT^{-1}s^{-1}$. For every held-out pair and
triple we therefore searched for a training program (single operation, trained pair, trained self-composition or
trained triple), a dihedral transform and a colour permutation under which the two compute the same function.
The colours an operation names (recolourings, swaps, borders and, in library B, painted rows, columns, corners
and diagonals) were found by testing its commutation with all colour transpositions. The colours a program names
are not always the colours that matter: in \emph{gravity left}, then \emph{recolour 9 to 3}, then \emph{recolour 9
to 1}, the last step is idle, because the first leaves no 9. Colour permutations were therefore enumerated as the
injective maps from the colours the training program names into the colours either program names, all of them
(at most 60\,480 for a program naming six colours), the identity first, and no candidate was rejected because the
two programs name different colours. Equality was tested on 19
grids built to separate operations (all cells distinct; all nine colours; non-square shapes) and 2\,000 fresh
random grids. Operations that break ties by colour index or by position (majority fill; in library B keeping the
most or least common colour, ranking colours, marking the densest row or column, sorting rows or columns) make a
conjugate differ from the original only where such a step meets a tie, so grids on which either program meets a
tie were left out and a match required at least 300 tie-free grids; candidates were screened only by output shape
and by which cells are empty, both unchanged by a colour permutation, and only on grids without a tie. Matches are
verified on these batteries, not proved, and equality is equality up to tie-breaking. A program is \emph{new} if
no training program computes it under any transform and permutation. On the main library 6 of 212 held-out pairs
and 100 of 300 held-out triples are new, and all six pairs occur as consecutive operations of a training triple,
which is why generalisation is tested on triples. Every library was audited against its own split (new triples:
49--100 of 300 in the libraries built from library A, 206--209 in library B), and every checkpoint was evaluated
on its library's new triples with ten episodes per triple and on its new pairs with twenty (Supplementary
Section~S1).

\subsection*{Measures}
\textbf{Solving.} Exact match on the whole grid, whose identity floor is zero on this library; cell
accuracy over cells that are non-padding in target or prediction; and accuracy on the cells the rule
changes. For RE-ARC we report exact match on queries whose output differs from the input and whose
target is not a single colour.

\textbf{The code.} Purity is the fraction of an operation's episodes that receive its modal message.
Distinct messages per rule is the number of distinct messages over the number of distinct rules in
the evaluation set, whose evaluation episodes cycle over the held-out rules. For RE-ARC, naming
statistics are computed over the rule families a run was trained on, with ten episodes per family.
\textbf{Decoding.} The episodes of each family are split in two halves; a decoder is fitted on one half and
scored on the other, and the two scores are averaged. Exact lookup assigns a test message the majority
family of identical messages in the fitted half (an unseen message counts as an error); nearest neighbour
assigns the family of the closest fitted message in Hamming distance, ties broken at random. The null
distribution shuffles family labels in the fitted half (200 permutations). Independent random messages
decode at chance (0.003 for 239 families, chance 0.004), whereas the plug-in normalised mutual information
between family and message reaches 1 on the same messages. Topographic similarity is the Spearman correlation between rule edit
distance and message Hamming distance; positional disentanglement follows Chaabouni et
al.~\cite{chaabouni2020}; name reuse is the fraction of held-out pairs whose message contains the
stage-one name of a component, with chance computed from an operation not in the pair.

\textbf{Information in the code.} Distinct messages count variation within a rule as well as between rules,
so a code is described by its names, the distinct modal messages over a set of rules, and by how often its
message identifies the rule on held-out episodes, with the decoders above: twenty episodes of each single
operation and ten of each new triple, generated by the run's own task generator and decoded from the speaker's
most probable message.

\textbf{Mechanism.} Lesions replace the symbol at one position with a different random symbol, or the whole
message with the message the speaker produced for another episode (Fig.~\ref{fig:emergence}d); with uniformly
random symbols instead, which the executor never met in training, cell accuracy on held-out pairs is 0.35
(gradient) and 0.19 (reward). Transmission trains a freshly initialised executor for 2\,000 steps on the frozen
speaker's messages. The co-adaptation test trains a new executor for 8\,000 steps (learning rate
$3\times10^{-4}$ after 200 warm-up steps, batch 32, single operations and trained pairs with the training
augmentation) on the frozen speaker's most probable messages, and in its blind control on the same messages
permuted across the episodes of each batch; it is compared with the original executor on the same episodes, ten
per new triple and 150 single operations. Serial execution is probed by decoding the executor's state after
each symbol and comparing it with the true intermediate grid of a two-operation rule, against the input and
output as references.

\textbf{Human iterated learning.} The chains of Kirby, Cornish and Smith \cite{kirby2008} were taken
from Edinburgh DataShare (doi 10.7488/ds/1586); for each chain and generation we computed the number
of distinct signals over the twenty-seven meanings, and topographic similarity between meaning
Hamming distance and signal edit distance. \textbf{Human descriptions.} For the comparison with
natural-language descriptions we used the LARC corpus \cite{acquaviva2022}: for each pair of ARC-1
tasks, the cosine similarity of the pooled descriptions (term frequency weighted by inverse document
frequency) and, for the 76 tasks with concept annotations, the cosine similarity of their concept
profiles; Mantel correlations with the same-message and Hamming-similarity matrices of the emergent
codes used 2\,000 label permutations, with and without residualisation on human build success and
description length.

\subsection*{A pretrained model in the same game}
Speaker and executor are two copies of Qwen2.5-0.5B-Instruct, each adapted by low-rank adapters (rank 16,
$\alpha=32$, on the attention and feed-forward projections; 8.8\,M trainable parameters of 503\,M). The speaker
reads the demonstrations as text and emits eight symbols sampled from a fixed alphabet of 32 tokens with no English
meaning; the executor reads the symbols and the query grid as text and writes the output. The reward, the group
baseline of four messages per episode, the entropy floor and the information term are those of the main
experiments; 5\,000 steps, batch 16, learning rate $10^{-4}$ with the speaker at 0.3 of it. The reward is the
accuracy of the output tokens under teacher forcing, where the main experiments reward the executor's cell accuracy;
both are evaluated by exact match of the whole grid. A generative executor almost never writes a whole correct grid
early in training. These episodes are not augmented, so the
held-out programs that count as new are those no training program computes. Every run is counted at each stage; runs whose
16-bit adapters became non-finite late in training are evaluated at their last snapshot before the event
(Supplementary Section~S9).

\subsection*{Statistics}
The unit of replication is an independent training run. Accuracy on held-out programs varies at two levels,
between runs and between programs, and every run of a library is evaluated on the same programs. Differences
between conditions are therefore reported with 95\% confidence intervals from a hierarchical bootstrap
\cite{saravanan2020} that resamples runs within each condition and, jointly for the conditions compared, the
held-out programs (10\,000 resamples, percentile intervals), with the two-sided bootstrap $p$-value. Welch's
$t$-test and an exact permutation test on run means are reported alongside; with two or three runs per condition
the smallest attainable permutation $p$-value is large, and it is given. Where a condition has a single run its
interval reflects variation over programs only. For decoding the resampled units are runs and rule families,
and for the human chains, chains. We treat three contrasts on new triples as primary, the ones on which the
main conclusions rest: staged against joint experience under reward, joint experience with against without the
information term, and the interaction between order of experience and channel. All other contrasts are
exploratory; no correction for multiple comparisons was applied, a difference whose interval includes zero is
not read as evidence of equality, and every contrast tested is listed in Supplementary Table~S11.

\subsection*{Compute and availability}
All runs used eight NVIDIA V100 GPUs; a curriculum run takes three to ten GPU-hours. Every number in
the paper is regenerated from the recorded evaluations by the analysis scripts; none is transcribed by hand.

\section*{Data availability}
The recorded evaluations of every training run reported, the outputs of every probe and audit, and the
per-episode messages used for decoding will be deposited in a public repository on publication. The human
iterated-learning chains are available from Edinburgh DataShare (doi 10.7488/ds/1586); LARC and the RE-ARC
generators are publicly available from their authors.

\section*{Code availability}
The code for the task generators, the networks, training, the audit of the compositional split, decoding,
and every figure and supplementary table will be released in a public repository on publication.

\subsection*{Figure conventions}
Individual runs are always shown; summaries never replace them. One colour is used for each channel
throughout: grey for the unconstrained real-valued channel, ochre for the gradient-trained discrete
channel, blue for the reward-trained one. Dashed lines mark the reference stated in each caption: the real-valued channel, or chance. Exact
match on the whole grid is the primary measure; its floor, the trivial predictor that copies its
input, is zero on the compositional libraries because every rule changes its input.

\section*{Acknowledgements}
The author acknowledges the CINECA award under the ISCRA initiative (ISCRA Class C project
 LUCID, HP10CLEJAD) for the availability of high-performance computing resources and support on Leonardo Booster,
 and the APE Lab, INFN, Sezione di Roma, for access to the server with eight NVIDIA V100 GPUs on which the
 experiments reported here were run. This work received no specific funding.

\section*{Author contributions}
C.C. conceived the study, designed and ran the experiments, analysed the data and wrote the manuscript.

\section*{Competing interests}
The author declares no competing interests.

\section*{Correspondence}
Correspondence should be addressed to C.C.~(cristiano.capone@iss.it).

\bibliography{refs}

\end{document}